\documentclass[preprint,amsmath,showpacs]{revtex4}
\usepackage{graphicx}
\usepackage{dcolumn}
\usepackage{bm}

\def\bee{\begin{eqnarray}}
\def\eee{\end{eqnarray}}

\begin{document}
\draft
\title{GZK Horizons and the Recent Pierre Auger Result on the
Anisotropy of Highest-energy Cosmic Ray Sources}
\author{Chia-Chun Lu} \affiliation{Institute of Physics, National
Chiao-Tung University, Hsinchu 300, Taiwan.}\author{Guey-Lin
Lin}\affiliation{Institute of Physics, National Chiao-Tung
University, Hsinchu 300, Taiwan}\affiliation{Leung Research Center
for Cosmology and Particle Astrophysics, National Taiwan University,
Taipei 106, Taiwan.}

\date{\today}

\begin{abstract}
Motivated by recent Pierre Auger result on the correlation of the
highest-energy cosmic rays with the nearby active galactic nuclei,
we explore possible ultrahigh energy cosmic ray (UHECR) source distributions and their effects on GZK horizons. Effects on GZK horizons by local over-density of UHECR sources are examined carefully with constraints on the degree of local over-density
inferred from the measured UHECR spectrum. We include the energy calibration effect on the Pierre Auger data in our studies. We propose possible local over-densities of UHECR sources which are testable in the future cosmic ray astronomy.
\end{abstract}


\maketitle

\section{Introduction}

Recently, Pierre Auger observatory published results on correlation
of the highest-energy cosmic rays with the positions of nearby
active galactic nuclei (AGN) \cite{Abraham:2007bb, AugerII}. Such a
correlation is confirmed by the data of Yakutsk \cite{Ivanov:2008it}
while it is not found in the analysis by HiRes \cite{Abbasi:2008md}.
In the Auger result, the correlation is maximal for the threshold
energy of cosmic rays at $5.7\times 10^{19}$ eV, the maximal
distance of AGN at $71$ Mpc and the maximal angular separation of
cosmic ray events at $\psi=3.2^{\circ}$. With the same threshold
energy, and the angular separation $\psi\leq 6^{\circ}$, the
correlation remains strong for a range of maximal AGN distance
between $50$ Mpc and $100$ Mpc. Due to increasing efforts on
verifying the Auger result, it is worthwhile to examine the above
correlation from a phenomenological point of view.

Since the angular scale
of the observed correlation is a few degrees, one expects that these cosmic ray particles are
predominantly light nuclei. The effect of GZK
attenuations on these cosmic ray particles \cite{Greisen,ZK} can be described by a distance scale referred to as ``GZK horizon" which is a function of the selected energy threshold for the arriving cosmic ray particles. By definition, the GZK horizon associated with a threshold energy $E_{\rm th}$ is the radius of a spherical region which is centered at the Earth and produce
$90\%$ of UHECR events arriving on Earth with energies above $E_{\rm
th}$. With continuous energy loss approximation, the GZK horizon for protons with
$E_{\rm th}=57$ EeV is about $200$ Mpc by assuming a uniformly
distributed UHECR sources with identical cosmic ray luminosity and
spectral index \cite{Harari:2006uy}. The calculations based upon
kinetic equation approach or stochastic energy loss also reach to
similar conclusions \cite{Kalashev:2007ph,Kachelriess:2007bp}.

The departure of theoretically calculated GZK horizon to the maximum
valid distance of the V-C catalog \cite{VC06} employed in Pierre
Auger's analysis, which is around $100$ Mpc, can be attributed to
several factors. As mentioned in \cite{AugerII}, such a deviation
may arise from non-uniformities of spatial distribution, intrinsic
luminosity and spectral index of local AGN. In addition, the energy
calibration also plays a crucial role since the GZK horizon is
highly sensitive to the threshold energy $E_{\rm th}$. Energy values
corresponding to the dip and the GZK cutoff of UHECR spectrum were
used to calibrate energy scales of different cosmic ray experiments
\cite{Berezinsky:2008qh,Kampert:2008pr}. It has been shown that all
measured UHECR energy spectra can be brought into good agreements by
suitably adjusting the energy scale of each experiment
\cite{Berezinsky:2008qh}. Keeping the HiRes energy scale unchanged,
the energy-adjustment factor $\lambda$ is found to be $1.2$, $0.75$,
$0.83$ and $0.625$ respectively for Auger, AGASA, Akeno and Yakutsk.
Furthermore, it has been shown that a different shower energy
reconstruction method infers a $30\%$ higher UHECR energy than that
determined by Auger's fluorescence detector-based shower
reconstruction \cite{Engel:2007cm}.

In this paper, we investigate the consistency between Auger's UHECR
correlation study and its spectrum measurement. As just stated, the
V-C catalog used by Pierre Auger for the correlation study is
complete only up to $100$ Mpc while the GZK horizon for $E_{\rm
th}=57$ EeV is generally of the order $200$ Mpc. We first consider
the local over-density of UHECR sources as a possible resolution to
the above discrepancy. It is motivated by the existence of Local
Supercluster (LS) which has a diameter of the order $60$ Mpc. In LS,
the over-density of galaxies has been estimated to be $\sim 2$
\cite{overdensity}.

The local over-density of UHECR sources has been invoked
\cite{Blanton:2000dr,Berezinsky:2001wn,Berezinsky:2002nc} to account
for AGASA data \cite{Hayashida:1994hb,Shinozaki:2004nh}. Such a
density distribution naturally leads to a smaller GZK horizon.
However, it also significantly affects the UHECR energy spectrum in
$(5\cdot 10^{19}-10^{20})$ eV region. Hence fittings to the measured
UHECR spectrum \cite{Roth:2007in} can provide information on the
degree of local over-density. Subsequently, the magnitude of GZK
horizon can be better estimated.

We next study the energy calibration effect on the estimation of
GZK horizon and the spectrum of UHECR. Certainly a $20\%-30\%$ upward shift on UHECR energies reduces the departure of theoretically calculated GZK
horizon to the maximum valid distance of V-C catalog \cite{AugerII}. The further implications of this  shift will be studied in fittings to the shifted Auger spectrum.

We fit the UHECR spectrum for events with energies above $10^{19}$
eV. This is the energy range where the GZK attenuation exhibits its
effect. It is also the energy range where the local over-density of
UHECR sources shows significant effects. In our analysis, we take
the UHECR as protons, which is hinted in the Auger events with
energies $\geq 57$ EeV although the composition study by the same
group suggests a heavier composition for $E\leq 40$ EeV
\cite{Unger:2007mc}. The HiRes experiment measures the composition
up to $50$ EeV \cite{Fedorova} and obtains a composition lighter
than that of Auger. For $E > 50$ EeV, the event number is still too
small for the composition study. To fit the UHECR spectrum at the
highest energy, it is more appropriate to treat the cosmic ray
energy loss as a stochastic process \cite{Stanev:2000fb}. There are
numerical packages available for treating stochastic energy loss of
cosmic ray particles \cite{Mucke:1999yb,Armengaud:2006fx}. We employ
the latter package for our calculations. Although UHECR loses its
energy mostly by scattering off CMB photons, it also loses some
amount of energy by scattering off infrared background photons
\cite{Stecker:1998ib,Franceschini:2001yg,Primack:2005rf,Lagache:2005sw,Stecker:2005qs}.
Thus we include the infrared photon contribution to the UHECR energy
attenuation. Source evolution $n(z)=n_0(1+z)^3$ is adopted in the
calculation of GZK horizon and spectrum, where $n_0$ is the source
number density at the present epoch.  It is from the
generally-accepted soft evolution model which traces the star
formation history and has been adopted in previous works
\cite{DeMarco:2005ia}.

We discuss about GZK horizons in Sec. II. We calculate the
accumulated event probabilities of UHECR for $E_{\rm th}=57$ EeV,
$70$ EeV, $80$ EeV and $90$ EeV respectively. GZK horizons
corresponding to different $E_{\rm th}$ are tabulated. We also
calculate GZK horizons with local over-density of UHECR sources
taken into account. In Sec. III, we fit the measured UHECR spectrum
with various local over-densities of UHECR sources and obtain information
on the degree of local over-density. To study the energy calibration effect, we also perform fittings to the shifted UHECR spectrum. Sec. IV contains
discussions and conclusions.

\section{The accumulative event probabilities of UHECR}
For single UHECR source, the cosmic-ray energy attenuation is
governed by the equation
\begin{equation}
\frac{\partial \phi_N(E,t)}{\partial t}=\frac{\partial}{\partial
E}\left[\left(-\frac{dE}{dt}\right)\phi_N(E,t)\right],
\label{conserve}
\end{equation}
in the continuous energy loss approximation. This equation results
from the number conservation of cosmic-ray particles in the energy
attenuation process. The cosmic-ray energy loss per unit time
$-dE/dt$ is due to the cosmic expansion and its scattering with
cosmic microwave background photons through photo-pion production
process $P\gamma \to N\pi$ and pair production process $P\gamma\to
Pe^+e^-$. The above attenuation equation is well known
\cite{kinetics}. In the current context, the solution of
Eq.~(\ref{conserve}) can be expressed in terms of the red-shift
variable \cite{Berezinsky:2002nc}
\begin{equation}
\phi_N(E,z)=\phi_N(\bar{E},z_s)\times
\exp\left[\int_{z}^{z_s}dz^{\prime}\left(\frac{(1+z^{\prime})}
{H(z^{\prime})}\times \frac{\partial b_0((1+z^{\prime})\bar{E})}{\partial \bar{E}}+\frac{1}{1+z^{\prime}}\right)\right],
\end{equation}
where $z_s$ is the red-shift of the UHECR source and the
function $b_0$ is related to the rate of cosmic-ray energy loss at
the present epoch by
\begin{equation}
-\frac{dE}{dt}(z=0)=b_0(E)+H_0E, \label{E_loss}
\end{equation}
where $H_0$ is the present value of Hubble constant. The UHECR has
an energy $\bar{E}$ at the source with red-shift $z_s$ and its
energy is downgraded to $E$ at the red-shift $z$. The energy
$\bar{E}$ is a function of $E$ and $z$ so that $\bar{E}(E,z_s)=E$
and
\begin{equation}
\frac{d\bar{E}}{dz}=-\frac{b_0\left((1+z)\bar{E}\right)}{H(z)}(1+z)-\frac{\bar{E}}{1+z}.
\label{Ebar_loss}
\end{equation}
Due to the non-trivial form of $b_0$, one resorts to numerical
methods for computing the function $\bar{E}$ and the flux
$\phi_N(E,z)$.

We have mentioned that the stochastic nature of UHECR energy loss
can not be overlooked for shorter propagation distances
\cite{Stanev:2000fb}. One then treats the energy attenuation by
photo-pion production as a stochastic process while treating other attenuations as continuous processes.
\begin{figure}[hbt]
\begin{center}
$\begin{array}{cc}
\includegraphics*[width=7.0cm]{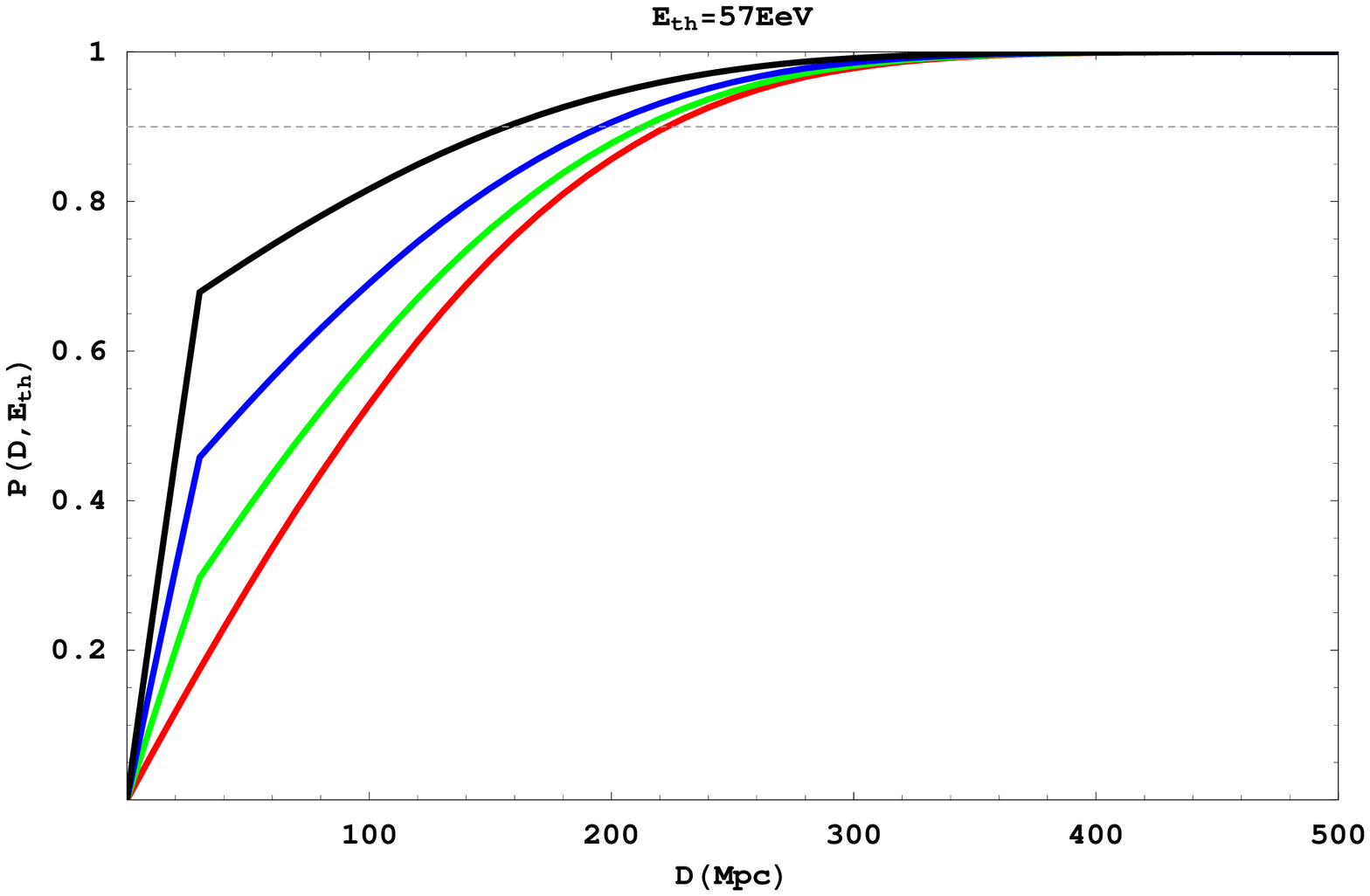} &
\includegraphics*[width=7.0cm]{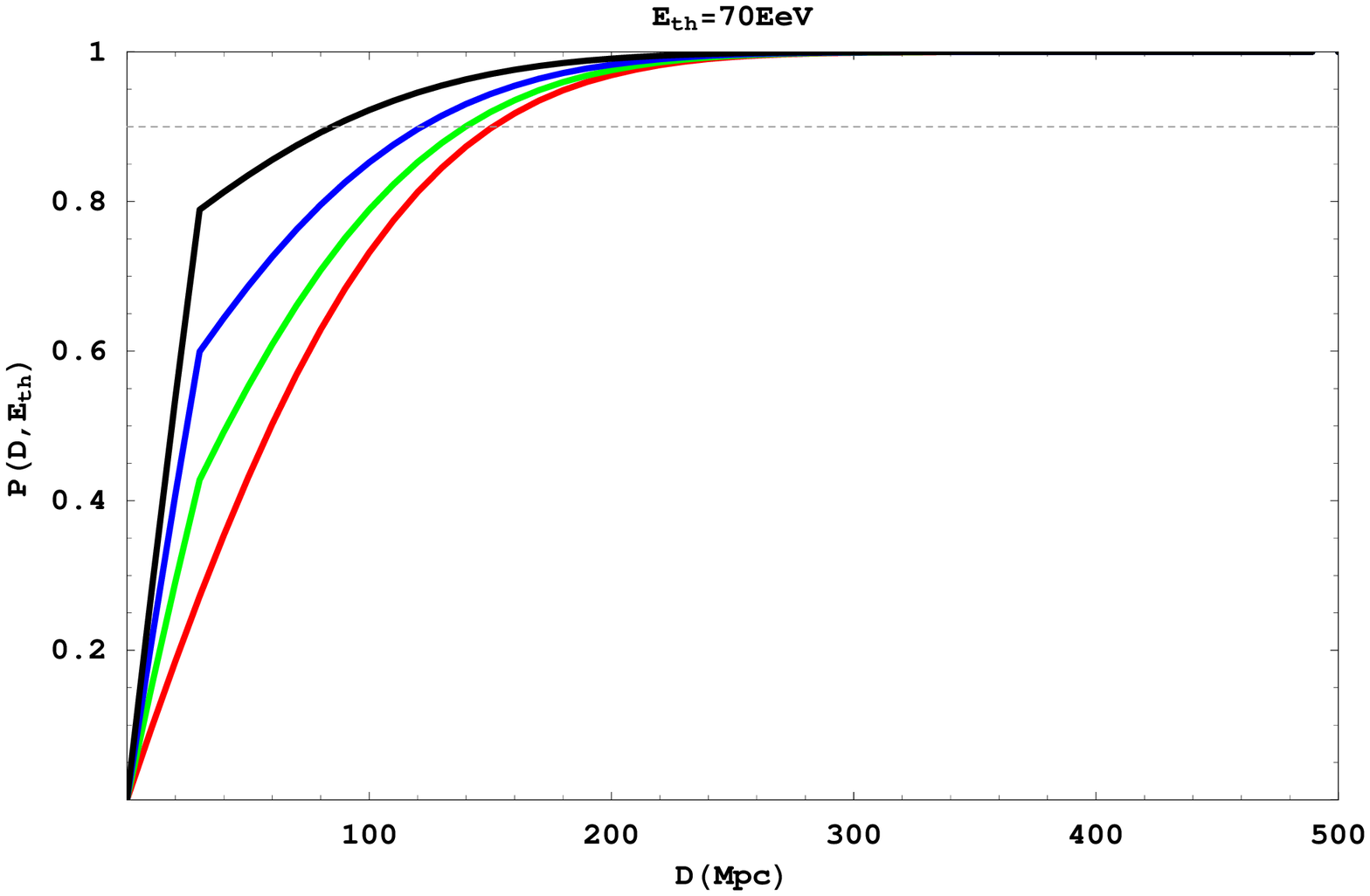}
\\
\includegraphics*[width=7.0cm]{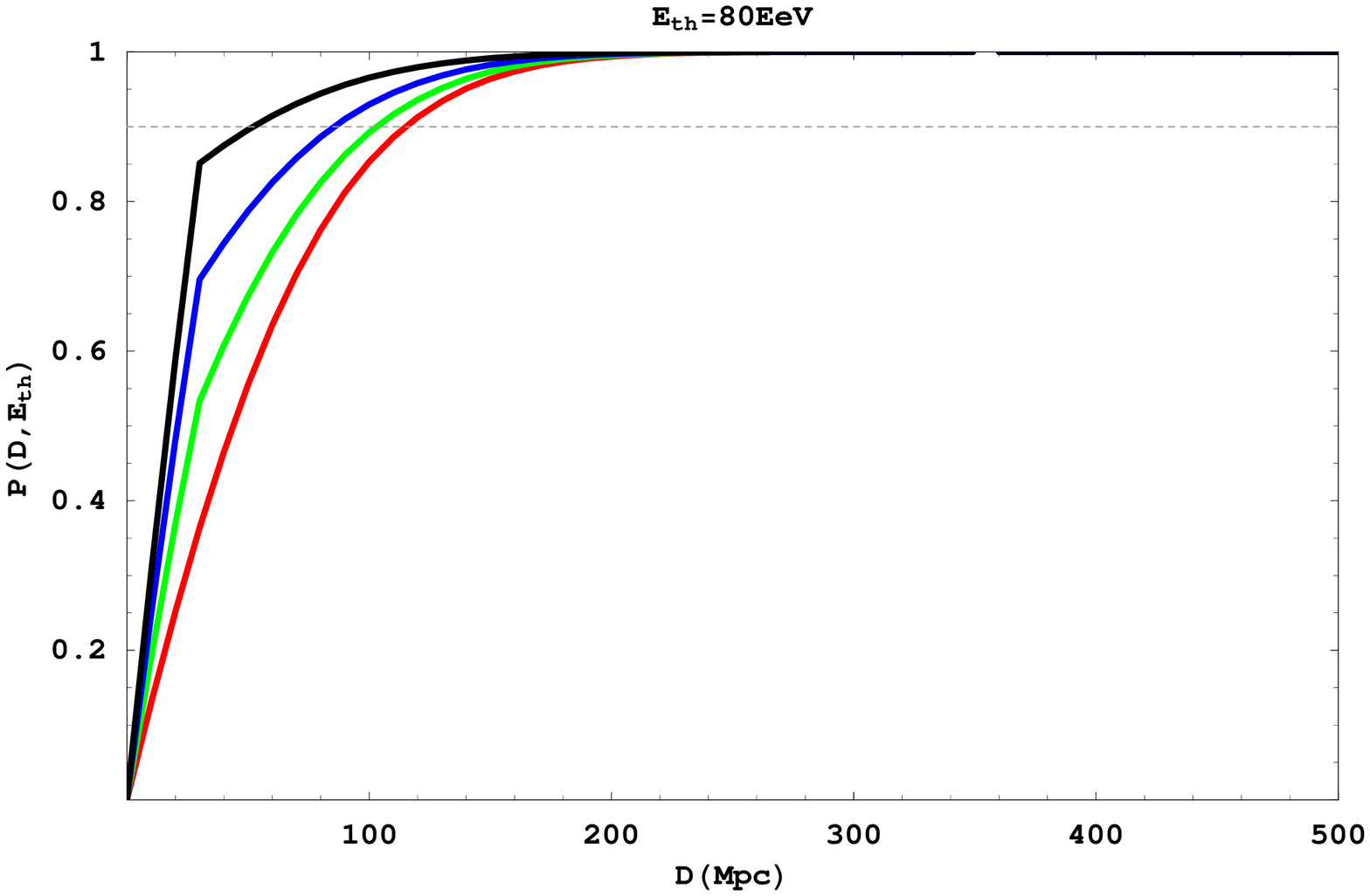} &
\includegraphics*[width=7.0cm]{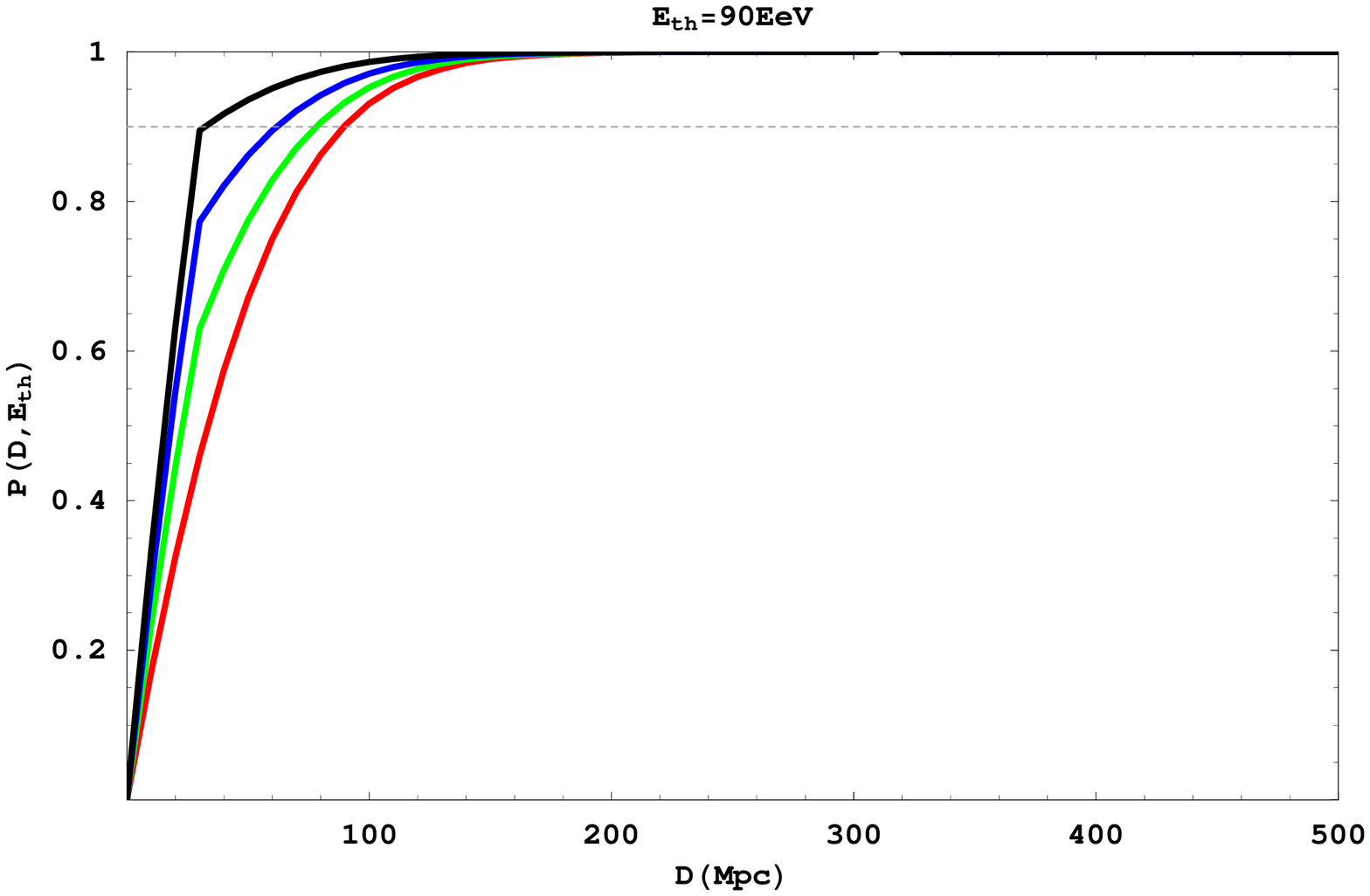}
\end{array}$
\end{center}
\caption{The accumulative event probability $P(D,E_{\rm th})$ as a
function of $D$ for $E_{\rm th}=57$ EeV, $70$ EeV, $80$ EeV and $90$
EeV respectively. The horizontal dash line in each panel denotes
$P(D,E_{\rm th})=0.9$. The red, green, blue and black curves
represent results from models with over-density $n(l<30 {\rm Mpc})/n_0=1, \ 2, \ 4,$ and $10$
respectively. The intrinsic spectrum index $\gamma=2.4$,
energy cut $E_{\rm cut}=1000$ EeV and the source evolution model
$n(z)=n_{0}(1+z)^3$ are used for calculations.}
\label{survival_prob}
\end{figure}

To facilitate our discussions, we define the accumulative event probabilities of UHECR as
\begin{equation}
P(D,E_{\rm th})=\frac{\int_{0}^{D}dl\cdot N(l,E_{\rm th})}{\int_{0}^{\infty}dl\cdot N(l,E_{\rm th})},
\end{equation}
where $N(l,E_{\rm th})\cdot dl$ is the number of cosmic ray events which are originated from sources at distances between $l$ and $l+dl$ from the Earth and arrive at the detector with energies above $E_{\rm th}$.
 We calculate $P(D,E_{\rm th})$ for
various local over-densities of UHECR sources. The source
distribution over the red-shift is taken as $n(z)=n_0(1+z)^3$ and
the energy spectrum of each source is taken to be the form,
$\phi_N(E)\equiv dN/dE=AE^{-\gamma}$, with the maximal energy
$E_{\rm cut}=1000$ EeV. We choose $\gamma=2.4$, $2.5$ and $2.6$
where $\gamma=2.5$ gives the best fitting to the measured UHECR
spectrum as will be shown in the next section. The accumulative
event probability $P(D, E_{\rm th})$ for $E_{\rm th}=57$ EeV, $70$
EeV, $80$ EeV and $90$ EeV are shown in Fig.~\ref{survival_prob} for
$\gamma=2.4$. Results for $\gamma=2.5$ and $\gamma=2.6$ are not
distinguishable from those for $\gamma=2.4$. In each panel, the red,
green, blue, and black curves represent local over-density $n(l<30
{\rm Mpc})/n_0=1, \ 2, \ 4,$ and $10$ respectively. The local
over-density $n(l<30 {\rm Mpc})/n_0=k$ is defined explicitly as
\begin{eqnarray}
n(l<30 {\rm Mpc})/n_0&=&k(1+z)^3,\nonumber \\
n(l\geq 30 {\rm Mpc})/n_0&=&(1+z)^3.
\label{over-density}
\end{eqnarray}
The horizontal dash line in each panel
denotes $P(D,E_{\rm th})=0.9$. The
intersection of this line with each color curve gives the GZK
horizon corresponding to a specific local over-density characterized
by the ratio $n(l<30 {\rm Mpc})/n_0$.
\begin{table}[hbt]
\begin{center}
\caption{GZK horizons of UHECR calculated with the local over-density $n(l<30 {\rm
Mpc})/n_0=1, \ 2, \ 4,$ and $10$, and arrival threshold energy
$E_{\rm th}=57$ EeV, $70$ EeV, $80$ EeV and $90$ EeV respectively. The listed numbers are in
units of Mpc.}
\begin{tabular}{|c|c|c|c|c|}\hline
$n(l<30 {\rm
Mpc})/n_0$ & $E_{\rm th}=57 \, {\rm EeV}$ & $E_{\rm th}=70 \, {\rm
EeV}$ & $E_{\rm th}=80 \, {\rm EeV}$ & $E_{\rm th}=90 \, {\rm EeV}$
\\ \hline $1$ & $220$ & $150$ & $115$ & $90$\\
\hline $2$ & $210$ & $140$ & $105$ & $75$\\
\hline $4$ & $195$ & $120$ & $85$ & $60$\\
\hline $10$ & $155$ & $85$ & $50$ & $30$\\
\hline
\end{tabular}
\end{center}
\label{horizon_p3}
\end{table}

GZK horizons corresponding to different local over-densities and
$E_{\rm th}$ are summarized in Table I. It is seen that local
over-densities up to $n(l<30 {\rm Mpc})/n_0=4$ do not alter GZK
horizons significantly for a given $E_{\rm th}$. One could consider
possibilities for higher local over-densities. However, there are no
evidences for such over-densities either from astronomical
observations \cite{overdensity} or from fittings to the measured
UHECR spectrum. We note that GZK horizons are rather sensitive to
$E_{\rm th}$. Table I shows that GZK horizons are $\sim 100$ Mpc or
less for $E_{\rm th}\geq 80$ EeV.

\section{Fittings to the UHECR spectrum measured by Pierre Auger}

As mentioned earlier, the local over-density of UHECR sources
affects the cosmic-ray spectrum at the highest energy, especially at
energies higher than $5\cdot 10^{19}$ eV. Hence the degree of local
over-density can be examined through fittings to the measured UHECR
spectrum as will be shown momentarily.

Fittings to the Auger spectrum have been performed in
\cite{Berezinsky:2005qe,Aloisio:2006wv,Arisaka:2007iz,Anchordoqui:2007fi}..
In our work, we take into account the over-density of UHECR sources
in the distance scale $l\leq 30$ Mpc. As stated previously, we take
the UHECR to be protons.
\begin{table}[hbt]
\begin{center}
\caption{The values of total $\chi^2$ from fittings to the Auger
measured UHECR spectrum. Numbers in the parenthesis are $\chi^2$
values from fittings to the $8$ data points in the energy range
$19.05\leq \log_{10}(E/{\rm eV}) \leq 19.75$. The last $4$ data
points record events with energy greater than $71$ EeV. }
\begin{tabular}{|c|c|c|c|c|}\hline
$n(l<30 {\rm
Mpc})/n_0$ & $1$ & $2$ & $4$ & $10$
\\ \hline $\gamma=2.5$ & $14.12 (9.34)$ & $14.61 (9.93)$ & $17.09 (10.50)$ & $28.09 (13.93)$\\
\hline $\gamma=2.6$ & $16.64 (12.28)$ & $15.56 (11.90)$ & $16.01 (11.83)$ & $20.76 (11.67)$\\
\hline
\end{tabular}
\end{center}
\label{chi_square}
\end{table}
\begin{figure}[hbt]
\begin{center}
$\begin{array}{c}
\includegraphics[width=9.5cm]{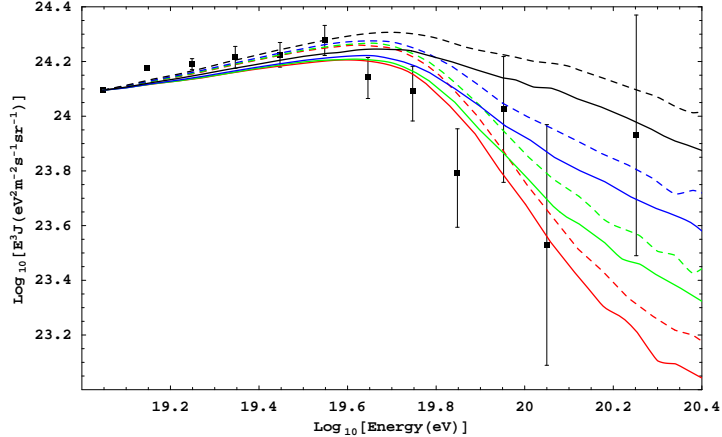}
\end{array}$
\end{center}
\caption{Fittings to the Auger measured UHECR spectrum where the
red, green, blue and black curves denote the model with the local
over-density $n(l<30 {\rm Mpc})/n_0=1, \ 2, \ 4,$ and $10$
respectively. Solid curves correspond to $\gamma=2.6$ while dash
curves correspond to $\gamma=2.5$. We take the source evolution
parameter $m=3$ throughout the calculations. } \label{unshift}
\end{figure}
Figure~\ref{unshift} shows our fittings to the Auger measured UHECR
spectrum with $\gamma=2.5$ and $2.6$ respectively. We take the
red-shift dependence of the source density as $n(z)=n_0(1+z)^m$ with
$m=3$. We have fitted $12$ Auger data points beginning at the energy
$10^{19}$ eV. We make a flux normalization at $10^{19}$ eV while
varying the power index $\gamma$ and the the degree of local
over-density, $n(l<30 {\rm Mpc})/n_0$. Part of $\chi^2$ values from
our fittings are summarized in Table II. We found that $\gamma=2.5,
\ n(l<30 {\rm Mpc})/n_0=1$ gives the smallest $\chi^2$ value with
$\chi^2/{\rm d.o.f.}=1.57$.
For the same power index, the large local over-density $n(l<30 {\rm
Mpc})/n_0=10$ is ruled out at the significance
level $\alpha=0.001$. For $\gamma=2.6$, $n(l<30 {\rm Mpc})/n_0=10$
is ruled out at the significance level
$\alpha=0.02$.

We note that, for both $\gamma=2.5$ and $\gamma=2.6$, the GZK
horizon with $n(l<30 {\rm Mpc})/n_0=10$, $E_{\rm th}=57$ EeV, $m=3$
and $E_{\rm cut}=1000$ EeV is about $155$ Mpc. Since $n(l<30 {\rm
Mpc})/n_0=10$ is clearly  disfavored by the spectrum fitting, one
expects a GZK horizon significantly larger than $155$ Mpc for
$E_{\rm th}=57$ EeV.

\begin{figure}[hbt]
\begin{center}
$\begin{array}{c}
\includegraphics[width=9.5cm]{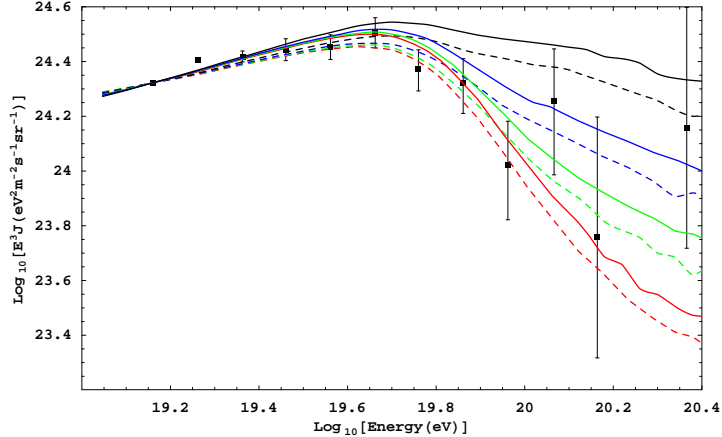}
\end{array}$
\end{center}
\caption{Fittings to the Auger measured UHECR spectrum with a
$30\%$ upward shift on UHECR energies where the red, green, blue
and black curves denote the model with the local over-density
$n(l<30 {\rm Mpc})/n_0=1, \ 2, \ 4,$ and $10$ respectively. Solid
curves correspond to $\gamma=2.4$ while dash curves correspond to
$\gamma=2.5$. We take the source evolution parameter $m=3$
throughout the calculations. } \label{shift}
\end{figure}

We next perform fittings to the shifted Auger spectrum. The results
are shown in Fig.~\ref{shift} where the cosmic ray energy is shifted
upward by $30\%$. Part of $\chi^2$ values are summarized in Table
III. The smallest $\chi^2$ value occurs approximately at
$\gamma=2.4$, $n(l<30 {\rm Mpc})/n_0=2$ with $\chi^2/{\rm d.o.f}=
0.82$. For $\gamma=2.5$, $\chi^2/{\rm d.o.f}=1.31$, $0.96$ and
$0.87$ for $n(l<30 {\rm Mpc})/n_0=1$, $2$ and $4$ respectively. It
is seen that $\chi^2$ values from current fittings are considerably
smaller than those from fittings to the unshifted spectrum. Given a
significance level $\alpha=0.1$, it is seen that every local
over-density listed in Table III except $n(l<30 {\rm Mpc})/n_0=10$
is consistent with the measured UHECR spectrum. It is intriguing to
test such local over-densities as will be discussed in the next
section.

We note that, with a $30\%$ upward shift of energies, the cosmic ray
events analyzed in Auger's correlation study would have energies
higher than $74$ EeV instead of $57$ EeV. The GZK horizon
corresponding to
$E_{\rm th}=74$ EeV is $120$ Mpc for $n(l<30 {\rm Mpc})/n_0=2$ and
$105$ Mpc for $n(l<30 {\rm Mpc})/n_0=4$.
\begin{table}[hbt]
\begin{center}
\caption{The total $\chi^2$ values from fittings to the Auger
measured UHECR spectrum with a $30\%$ upward shift on UHECR
energies. Numbers in the parenthesis are $\chi^2$ values from
fittings to the $8$ data points in the energy range $19.16\leq
\log_{10}(E/{\rm eV}) \leq 19.86$. The last $4$ data points record
events with energy greater than $92$ EeV. }
\begin{tabular}{|c|c|c|c|c|}\hline
$n(l<30 {\rm
Mpc})/n_0$ & $1$ & $2$ & $4$ & $10$
\\ \hline $\gamma=2.4$ & $8.65 (4.30)$ & $7.39 (4.67)$ & $10.26 (6.35)$ & $27.31 (13.34)$\\
\hline $\gamma=2.5$ & $ 11.82 (6.16)$ & $8.67 (5.49)$ & $7.78 (5.23)$ & $16.18 (7.39)$\\
\hline
\end{tabular}
\end{center}
\label{chi_square2}
\end{table}

We have so far confined our discussions at $m=3$. In the literature,
$m$ has been taken as any number between $0$ and $5$. It is
demonstrated that the effect on UHECR spectrum caused by varying $m$
can be compensated by suitably adjusting the power index $\gamma$
\cite{DeMarco:2005ia}. Since GZK horizons are not sensitive to
$\gamma$ and $m$, results from the above analysis also hold for
other $m$'s.

\section{Discussions and conclusions}

We have investigated the consistency between Auger's latest result
on the correlation of UHECR sources with positions of nearby
extra-galactic AGN and its measured UHECR spectrum. As stated
before, this investigation is motivated by the fact that the V-C
catalog used by Pierre Auger for the correlation study is reliable
only up to $100$ Mpc while the GZK horizon for $E_{\rm th}=57$ EeV
is generally of the order $200$ Mpc. We have explored the
possibility for local over-density of UHECR sources, which is
expected to shorten the GZK horizon for a given threshold energy of
arrival cosmic-ray particles. This is indeed the case as can be seen
from Table I. On the other hand, the effect is far from sufficient
to shorten the GZK horizon at $E_{\rm th}=57$ EeV to $\sim 100$ Mpc
for a local over-density of UHECR sources consistent with the
measured UHECR spectrum.

We have performed a upward energy shift to the Auger measured UHECR
spectrum. As said, a upward energy shift is motivated by simulations
of shower energy reconstructions as well as the requirement of
reproducing the theoretically predicted GZK cutoff energy. With a
$30\%$ energy shift, each cosmic ray event used by Auger for the
correlation study would have an energy above $74$ EeV instead of
$57$ EeV. GZK horizons corresponding to $E_{\rm th}=74$ EeV then
match well with the maximum valid distance of V-C catalog. Fittings
to the shifted Auger spectrum indicate a possibility for the local
over-density of UHECR sources.

We point out that the local over-density of UHECR sources is
testable in the future cosmic ray astronomy where directions and
distances of UHECR sources can be determined. Table IV shows
percentages of cosmic ray events that come from sources within $30$
Mpc for different values of $E_{\rm th}$ and $n(l<30 {\rm
Mpc})/n_0$. We take $\gamma=2.4$, $m=3$ and $E_{\rm cut}=1000$ EeV
for calculating these percentages. We note that these percentages
are not sensitive to the above parameters.
\begin{table}[hbt]
\begin{center}
\caption{ Percentages of cosmic ray events that come from sources
within 30 Mpc for different values of $E_{\rm th}$ and local over-density $n(l<30 {\rm
Mpc})/n_0$.}
\begin{tabular}{|c|c|c|c|c|}\hline
$n(l<30 {\rm
Mpc})/n_0$ & $E_{\rm th}=57 \, {\rm EeV}$ & $E_{\rm th}=70 \, {\rm
EeV}$ & $E_{\rm th}=80 \, {\rm EeV}$ & $E_{\rm th}=90 \, {\rm EeV}$
\\ \hline $1$ & $0.17$ & $0.27$ & $0.36$ & $0.46$\\
\hline $2$ & $0.30$ & $0.43$ & $0.53$ & $0.63$\\
\hline $4$ & $0.46$ & $0.60$ & $0.70$ & $0.77$\\
\hline $10$ & $0.68$ & $0.79$ & $0.85$ & $0.89$\\
\hline
\end{tabular}
\end{center}
\label{event_dist}
\end{table}
For $E_{\rm th}=57$ EeV, only $17\%$ of cosmic ray events come from sources less than $30$ Mpc away for $n(l<30 {\rm
Mpc})/n_0=1$. For $n(l<30 {\rm
Mpc})/n_0=2$ and the same threshold energy, $30\%$ of cosmic ray events are originated from sources in the same region.

It should be stressed that we have focused only on resolving the
apparent discrepancy between the GZK horizon at $E_{\rm th}=57$ EeV
and the maximum valid distance of V-C catalog. The statistics
analysis for establishing the source correlation is an independent
issue beyond the scope of the current paper. We have found that the
above discrepancy can not be resolved by merely introducing the
local over-density of UHECR sources. On the other hand, if Auger's
energy calibration indeed underestimates the UHECR energy, such a
discrepancy can be reduced. More importantly, fittings to the
shifted Auger spectrum indicate a possible local over-density of
UHECR sources, which is testable in the future cosmic ray astronomy.

\noindent{\bf Acknowledgements}

We like to thank A. Huang and K. Reil for helpful discussions. We
also thank F.-Y. Chang, T.-C. Liu and Y.-S. Yeh for assistances in
computing. This work is supported by National Science Council of
Taiwan under the grant number 96-2112-M-009-023-MY3.

\end{document}